\documentclass[pra,showpacs,showkeys, twocolumn,sort&compress, floatfix, amsfonts,square]{revtex4}

\usepackage[dvips]{graphicx}
\usepackage{bm}

\renewcommand{\deg}{$^{\circ}$}

\begin{document}

\title{The Stoner-Wohlfarth model of Ferromagnetism: Dynamic and Statistical properties}

\author{C. Tannous and J. Gieraltowski}
\affiliation{Laboratoire de Magn\'etisme de Bretagne - CNRS FRE 2697\\
Universit\'e de Bretagne Occidentale -\\
6, Avenue le Gorgeu C.S.93837 - 29238 Brest Cedex 3 - FRANCE}

\begin{abstract}
The physics of magnetic state change in single domain magnetic grains
 (called Stoner particles)
is interesting from the fundamental as well as the applied points of view.
A change in magnetization can be finely tuned with a specific time variation of 
an externally applied magnetic field.
It may also occur naturally (without application of a field) at very low temperature
with quantum tunneling and at higher temperature
with thermal excitation. The optimal (usually shortest) time altering the magnetisation along with 
the smallest applied magnetic field are sought in technological applications such
as high-density reading or writing of information, spintronics, quantum information and 
quantum communication systems. \\
This work reviews the magnetization change with a time dependent field and temperature
and discusses the time it takes to alter the magnetization as a function of 
the control parameter chosen, temperature and material parameters.
\end{abstract}



\maketitle

\section{Introduction} 
The effect of time dependent fields on magnetization state is important for the reading and
writing of information and the monitoring of the magnetization in a magnetic material.
In the case of magnetic recording, when density is increased, the grain size making the 
recording media decreases. If it is small enough, its magnetization becomes extremely sensitive to thermal 
energy; it can flip (in the vertical case) or reverse (in the horizontal case)
 by the simple effect of (even small) finite temperature perturbation
(Brownian fluctuations). This adverse effect is called super-paramagnetism
that traditionally limits longitudinal recording of hard disks to densities 
on the order of 100 Gbits/in$^{2}$. Longitudinal refers to the fact the rotation
velocity of the disk is parallel to the magnetization orientation. In perpendicular
recording this limit is ten times higher around several Tbits/in$^{2}$. \\

Progress in size reduction toward the nanometer paves the way to new opportunities in the
emerging  field of spintronics \cite{Zutic}. On that scale, we have a wide  panel of physical effects 
(e.g. new types of quantum exchange between nanometer thick magnetic layers) 
and the spin diffusion length becomes long enough to maintain useful spin orientation. 
Novel nanometric magnetic devices are good candidates for use as building blocks of 
spintronics (spin diode or spin transistor) or quantum information systems. The latter span
quantum information storage (Q-bits), quantum computing (quantum logic operations like the 
square root of the $NOT$ operation ($\surd NOT$) or the controlled $NOT$ operation ($CNOT$), the
Quantum Fast Fourier Transform...), 
quantum communication systems (an example is entanglement which means that measurement performed on one system
seems to be instantaneously influencing other systems related to it) or Quantum Metrology. \\

It is important to be able to tell how
one might be able to alter some state with a magnetic field, or how it might be affected
by temperature as in the recording case. 
It is important to point out that interesting quantum phenomena
might occur at low temperatures ($T \sim 0K$) such as magnetic quantum tunneling; while this is 
beyond the scope of this paper, the reader might consult a review such as  ref.~\cite{Chud}. \\

The time it takes for an effect to take place is also important.
In recording, given a fixed rotational velocity of the hard disk 
(typically 7200 rpm) the decrease of bit length, imposes a faster (higher frequency)
process of reading (sensing the magnetization orientation)/writing (changing the
magnetization orientation) of the bit. While the shortest
time altering the magnetization is required in reading/writing  applications, the longest time
is required in (long-term) storage with protection of the data against large  magnetic
fields that might corrupt or even erase the stored information. \\
 
In this paper corresponding to the second part of the series on the Stoner-Wohlfarth model, 
we examine the effects of time dependent field and finite temperature on a 
single domain Stoner particle \cite{Stoner}. The time it takes for the magnetization to change is also studied 
with temperature and material parameters. \\

This paper is organized as follows: in section 2 we examine the evolution of magnetization
state with a time dependent field. In section 3 we discuss the effect of temperature 
on magnetization reversal and we conclude in section 4 with the possible extensions and
perspectives of the SW model.

\section{Equation of motion for the magnetization in the presence of a time dependent field}
Magnetization dynamics is governed by the Landau-Lifshitz-Gilbert equation. Since
magnetization  $\bm{M}$ is akin to angular momentum, we have an evolution equation for  $\bm{M}$
similar to angular momentum, the Bloch equation of motion  $d\bm{M}/dt=\gamma_0 \bm{M} \times \bm{H}$
with $\gamma_0$ the gyromagnetic ratio and $\bm{H}$ the external field (see for instance Kittel \cite{Kittel}).

Extending the Bloch equation to a  moment $\bm{M}$ subjected  to an "effective" field  $\bm{H_e}$
and a dissipation term describing  losses and relaxation processes in the material, 
Landau and Lifshitz (L-L) \cite{Landau} assumed that
dissipation is accounted for by a coefficient $\lambda$ and introduced a dissipation 
non-linear term of the form  $\frac{\lambda \gamma_0}{\|\bm{M}\|} \bm{M} \times (\bm{M} \times \bm{H_e}) $
(where $\|\bm{M}\|$ is the modulus of $\bm{M}$) controlled by the effective magnetic field $\bm{H_e}$:

\begin{equation}
\frac{d\bm{M}}{dt} =  -\gamma_0 (\bm{M}\times \bm{H_e})
+\frac{\lambda \gamma_0}{\|\bm{M}\|} \bm{M} \times (\bm{M} \times \bm{H_e})
\label{eq:LL1}
\end{equation}

L-L define, as in Quantum Field Theory, the effective field $\bm{H_e}$  from the 
functional derivative of the total energy
with respect to magnetization $\bm{H_e}=-\delta E/\delta \bm{M}$; hence in any magnetic 
problem the total energy should be the starting point whether one is dealing with static or dynamical
problems. In the simple Stoner particle case, the functional derivative reduces to the gradient 
with respect to  $\bm{M}$, $\bm{H_e}=-\partial E/\partial \bm{M}$. 

In order to avoid the divergence problem arising in the L-L equation for the 
large dissipation case ($\lambda >> 1$), Gilbert modified the L-L dissipation term by introducing a 
damping term of the form $\alpha \gamma  \bm{M} \times (\frac {d\bm{M}}{dt})$.

The equation of motion of a magnetic moment in presence of damping and effective
field is given by the Landau-Lifshitz-Gilbert (L-L-G) equation:

\begin{equation}
 \frac{d\bm{M}}{dt} =  -\gamma (\bm{M}\times \bm{H_e})
+ \alpha \gamma  \bm{M} \times (\frac {d\bm{M}}{dt})
\label{eq:LLG}
\end{equation}

where $\bm{M}$ is the magnetization vector, $\bm{H_e}$ the effective field, 
$\gamma$ another gyromagnetic ratio and $\alpha $ the damping parameter. 

First of all, we retrieve Bloch equation in the simple case of zero damping and
effective field $\bm{H_e}=\bm{H}$ the externally applied field.
In the static case, the L-L-G equation reduces to  $\bm{M}\times \bm{H_e}=0$ meaning
the static equilibrium condition is either $\bm{M} // \bm{H_e}$ or 
$\bm{H_e}=0$ equivalent to the extremum (minimum)
condition on the energy as discussed previously (see first part of this work). \\
The L-L-G equation conserves $\|\bm{M}\|=\sqrt{\bm{M} \cdot \bm{M}}$ as seen by
taking the scalar product on both terms of the RHS of eq.~\ref{eq:LLG}.  One gets 
$ \frac{d\bm{M}}{dt}\cdot \bm{M}=0$ meaning that $\|\bm{M}\|= M_s=$ constant, where
$M_s$ is the saturation magnetization.\\
The L-L-G equation seems odd from the mathematical point of view since one is
used (in systems of ordinary differential equations or ODE) to see the first
derivative term $\frac{d\bm{M}}{dt} $ in the LHS only. Here it appears on both sides
and pushes one to think that the system cannot be handled by standard mathematical
integration tools like Euler or Runge-Kutta  methods. \\
In addition, it is misleading to  attempt at solving recursively the L-L-G equation 
by  substituting repeatedly the term  $ \frac{d\bm{M}}{dt}$ in the RHS of the equation.
It is straightforward to show that the Landau-Lifshitz (L-L) 
equation is mathematically  equivalent
to the L-L-G equation by taking the cross product of the LHS of eq.~\ref{eq:LLG} with $\bm{M}$
and using $\bm{M}$ norm conservation ($ \frac{d\bm{M}}{dt}\cdot \bm{M}=0$).
We obtain:

\begin{equation}
\bm{M} \times  \frac{d\bm{M}}{dt} = - \gamma \bm{M} \times (\bm{M}\times \bm{H_e})
+ \alpha \gamma   M_s^2 (\frac {d\bm{M}}{dt})
\end{equation}

Substituting $\bm{M} \times  \frac{d\bm{M}}{dt}$ in eq.~\ref{eq:LLG}, we get:

\begin{eqnarray}
\frac{d\bm{M}}{dt} = -\frac{ \gamma}{1+\alpha^2}  (\bm{M}\times \bm{H_e}) \hspace{3cm} \nonumber   \\
\hspace{3cm}  + \frac{\alpha \gamma}{1+\alpha^2} \bm{M} \times (\bm{M} \times \bm{H_e})
\label{eq:LL2}
\end{eqnarray}

It is now a matter of interpreting the coefficients appearing in the 
L-L  or the L-L-G equations that will make them differ in a given situation.
If one identifies $\gamma_0$  as $\frac{ \gamma}{1+\alpha^2}$ and 
$\lambda \gamma_0$ as $\frac{\alpha \gamma}{1+\alpha^2}$ then both equations are 
same but if one insists on keeping $\gamma_0$ as the gyromagnetic ratio or
confusing dissipation ($\lambda$) and damping  ($\alpha$) then
the equations will differ since the factors affecting both terms in the RHS are 
numerically different. \\
In addition, the L-L dissipation term goes to zero when the damping coefficient
goes to infinity making the L-L-G equation appear more physically appealing than the L-L equation.

The system of ODE eqs.~\ref{eq:LL2} is integrable 
by standard explicit methods, such as Euler or Runge-Kutta
(see for instance ref.~\cite{Recipes}) after expressing the components in  Cartesian 
coordinates. The conservation
of the norm is very useful during integration (specially in explicit integration 
schemes) to test the accuracy and stability of integration.

\begin{eqnarray}
\pmatrix{\dot{m}_x \cr \dot{m}_y \cr \dot{m}_z}
=-\frac{\gamma H_K}{(1+\alpha^2)} \times  \hspace{4cm}   \nonumber \\
\pmatrix{(1+\delta_x^2) & -(\delta_z-\delta_x\delta_y) & (\delta_x\delta_z+\delta_y) \cr
         (\delta_z+\delta_x\delta_y) & (1+\delta_y^2) & -(\delta_x-\delta_y\delta_z) \cr 
(\delta_x\delta_z-\delta_y) & (\delta_x+\delta_y\delta_z) & (1+\delta_z^2)}
\times   \nonumber \\
 \pmatrix{m_y h_{ez} - m_z h_{ey} \cr m_z h_{ex} - m_x h_{ez} \cr m_x h_{ey} - m_y h_{ex} }  \hspace{1cm}
\end{eqnarray}

with the definitions: $\bm{m}=\bm{M}/M_s$, $\dot{\bm{m}}=\frac{d\bm{m}}{dt} $,
$\bm{h_e}=\bm{H_e}/H_K$ and $\delta_x=\alpha(M_x/M_s), \delta_y=\alpha(M_y/M_s),\delta_z=\alpha(M_z/M_s)$.

\begin{figure}[!ht]
\begin{center}
\scalebox{0.8}{\includegraphics*[angle=0]{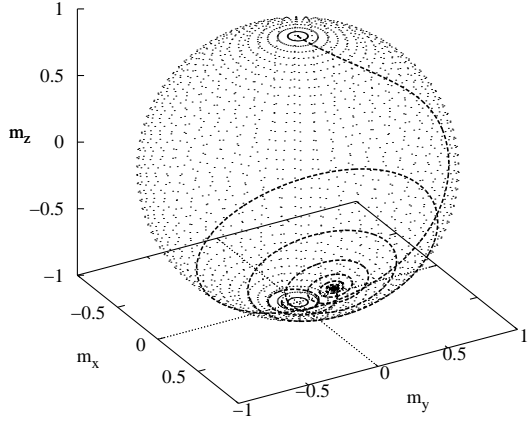}}
\end{center}
  \caption{Trajectory of the magnetization tip on the unit sphere for a field 
applied at t=0  in the y-z plane and making an angle of 135 \deg with the z-axis. The damping
is $\alpha=0.1$, the field is in the yOz plane making an angle of 135 \deg with z-axis. It is applied
at $t=0$ for 9 nanosecs. Its value is 0.5 $H_K$.}
\label{traj}
\end{figure}

\begin{figure}[!ht]
\begin{center}
\scalebox{0.5}{\includegraphics*[angle=0]{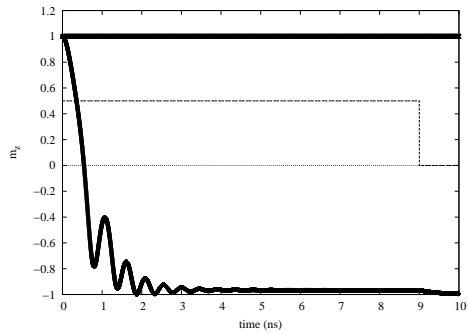}}
\end{center}
  \caption{Variation of the magnetization component $m_z$ as a function of time.
   The parameters are the same as in fig.~\ref{traj}.
    The straight thick line indicates conservation of $\|\bm{M}\|$ during integration. The
dotted line is the variation of the applied magnetic field with time. The ringing observed
due to damping is a major cause of delay in reversal.}
\label{mz}
\end{figure}

Using  order-4 Runge-Kutta (RK4) method (see ref.~\cite{Recipes}) with $\bm{M}$
along the z-axis as an initial condition, we apply at $t=0$ a time dependent field making 135\deg with the z-axis
(see first part of this work).
The 3D response of the magnetization in time is depicted in fig.~\ref{traj} and the z-component
of $\bm{m}$ is depicted in fig.~\ref{mz}. Undesirable ringing effects (oscillations)
in the time variation of $\bm{m}$ are observed. They are so because they
introduce an unwanted  delay in magnetization reversal.

\begin{figure}[!ht]
\begin{center}
\scalebox{0.8}{\includegraphics*[angle=0]{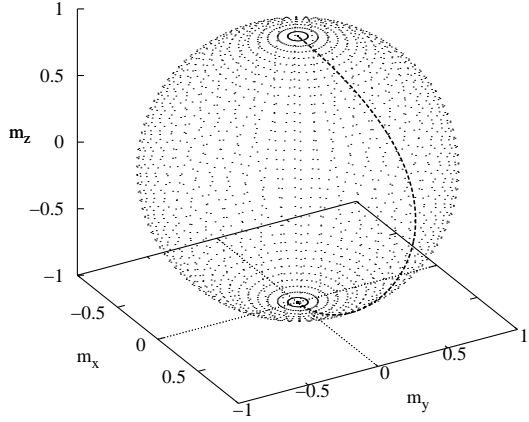}}
\end{center}
  \caption{Ballistic trajectory of the magnetization tip on the unit sphere for a field 
applied at t=0  in the y-z plane and making an angle of 90 \deg with the z-axis. The damping
is small: $\alpha=0.001$, the field is in the yOz plane making an angle of 90 \deg with z-axis. It is applied
at $t=0$ for 0.12 nanosecs. Its value is 1.7 $H_K$.}
\label{ballistic}
\end{figure}

\begin{figure}[!ht]
\begin{center}
\scalebox{0.5}{\includegraphics*[angle=0]{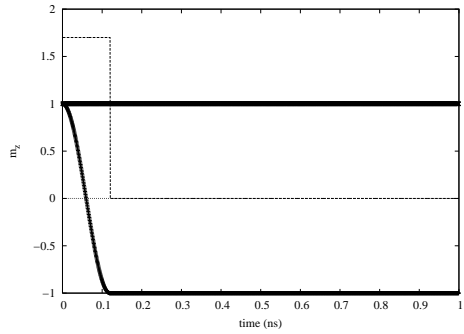}}
\end{center}
  \caption{Variation of the magnetization component $m_z$ as a function of time. 
  The parameters are the same as in fig.~\ref{ballistic}.  
  The straight thick line indicates conservation of $\|\bm{M}\|$ during integration. The
dotted line is the variation of the applied magnetic field with time. No delay in magnetization
reversal is observed due to the small damping suppressing the ringing.}
\label{mz_bal}
\end{figure}

In order to eliminate the ringing effect, we move on to another reversal mode called precession
switching in which the field is applied perpendicularly to the initial magnetization and
whose action is on until the magnetization is reversed without displaying any ringing effect.
The reversal path on the unit sphere is called
a ballistic path (see fig.~\ref{ballistic}) emphasizing its optimality. 
The sensitivity of this process stems from the fact,
the field must be switched off exactly at the time magnetization reverses (see fig.\ref{mz_bal}).

\section{Effect of temperature on magnetization dynamics}

\begin{figure}[!ht]
\begin{center}
\scalebox{0.3}{\includegraphics*[angle=0]{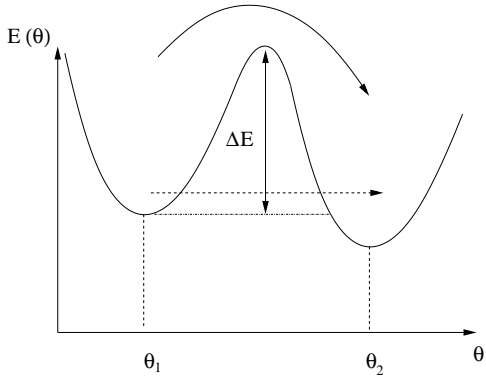}}
\end{center}
  \caption{Energy versus angle $\theta$ showing the barrier $\Delta E$ the system
 has to overcome in order to go from state (1) with $\theta_1$ to  state (2) with $\theta_2$. At low
temperature the system can tunnel from state (1) to  state (2).}
\label{transition}
\end{figure}

A grain at finite temperature is prone to thermal excitations that might
alter its magnetization state. The simplest model describing the effect of
temperature on a grain is inspired from Chemistry and is called the N\'eel-Arrhenius
thermal excitation model. \\
At very low temperature, switching may occur by tunneling  at a given energy 
through the energy barrier  separating two magnetization states corresponding 
to two energy minima (see fig.~\ref{transition}). This is known as Macroscopic
Quantum Tunneling of Magnetization that we will not describe here but for which there
exist many reviews  (see for instance ref.~\cite{Chud}).

At finite temperature, the empirical Arrhenius model is used to describe the kinetics of 
a thermally activated process. 
This assumes that an energy barrier hinders the forward progress of a chemical reaction.
The height of this energy barrier is a measure of resistance to the reaction.
Forward progress of the reaction requires the supply of an activation energy to surmount
this barrier. It has the form \cite{Condon}: \\
\begin{equation}
\tau=\tau_0 \exp(\Delta E/k_B T)
\label{barrier}
\end{equation}
where $\tau$ is the chemical reaction "inverse rate", $\tau_0$ is the attempt time to traverse the barrier,
$\Delta E$ is the barrier height, $k_B$ is Boltzmann constant and $T$ is absolute temperature.

Drawing an analogy from radioactivity, one might view switching as a decay process 
with a typical probability of decay $\lambda$. Starting from an assembly of grains
$N_0$ at $t=0$, the number of particles that decay in the instant $[t,t+dt]$ is
$dN=-\lambda N(t) dt$. Integrating  with the initial condition $N_0=N(t=0)$ we find that
the particles that are still present (did not decay or switch) is given
by $N(t)=N_0 \exp(-\lambda t)$. This analogy holds if switching is treated as an
irreversible process like decay. This means, switching back to the original
value is not considered as a valid process. This is the case of data recording: If
the stored value has changed once, it is no longer valid and must be rejected.\\

Since the average lifetime is given by $\tau=1/\lambda$ we interpret the inverse rate as the
average lifetime with respect to switching. This means that the recorded information in
a magnetic material (tape, hard disk, floppy etc...) stays unaltered for a period of time given by $\tau$.
We infer from this analogy that the probability of switching is given by $\exp(-t/\tau)$ and 
therefore the probability of retaining the information (not switching) is given by
the complementary probability: $P(t)=1-\exp(-t/\tau)$ with the new interpretation of Arrhenius
formula $\tau=\tau_0 \exp(\Delta E/k_B T)$.

This decay picture of switching can be recast in a two-level model since switching means
we have a transition from a magnetization state (1) to another (2) as depicted in fig.~\ref{transition}.

Considering a number (normalized) of non-interacting grains in state (1) as $n_1$ and
the number of grains in state (2) as $n_2$ we may write a kinetic equation (Master
equation) with typical transition times $\tau_1, \tau_2$ as:

\begin{equation}
\frac{dn_1}{dt}=\frac{n_1}{\tau_1}-\frac{n_2}{\tau_2}
\end{equation}

Assuming total number (normalized) conservation: $n_{1}+ n_{2}=1$, 
the solution of this equation  is given by:

\begin{equation}
n_{1,2}=\frac{\tau_{1,2}}{\tau_1+\tau_2} \pm [n_{1,0}- \frac{\tau_{1}}{\tau_1+\tau_2} ] \exp(-t/\tau) 
\end{equation}
 
where $n_{1,0}$ is the initial value of $n_{1}$ i.e. $n_{1,0}= n_1(t=0)$.
It is interesting to note that the decay time $\tau=\frac{\tau_{1} \tau_{2}}{\tau_1+\tau_2}$ is the
geometric average of $\tau_1   \mbox{   and    }   \tau_2$.

As a result, we obtain a simple classification of the possible magnetic states:
\begin{itemize}
\item We have a blocked state when $\tau >> t$ i.e. $n_{1}= n_{1,0} \mbox{     }  \forall t$.
\item We have a super-paramagnetic state in the opposite case $\tau << t$ leading to
$n_{1,2}=\frac{\tau_{1,2}}{\tau_1+\tau_2}$.
\end{itemize}

Physically, $t$ is of the order of the experimental measurement time and a blocked state
means that no change to the system is observed during $t$. On the other hand, when the instrinsic 
time $\tau << t$, the magnetization change is so frequent that no 
well defined state is maintained for a long enough time.
Thus the system behaves like a paramagnetic system that cannot store 
information (in a stable and reliable way). Hence the origin of the "super-paramagnetic" qualifier.

When a grain switches we have information storage errors and the bit error rate (BER) is given by
the switching probability $\exp(-t/\tau)$. 

In order to appreciate the meaning of BER and therefore average lifetime and 
barrier height, suppose we impose a BER of 10$^{-12}$. This means one bit is wrong in a 
hard disk of 125 GBytes capacity. Identification of BER and 
switching probability $\exp(-t/\tau)$ means that $t=10^{-12} \times \tau$.
According to eq.~\ref{barrier} and with the assumptions: $\tau_0 \sim 10^{-9}$ sec 
and $\Delta E/k_B T=68$ we get $t \sim \pi \times 10^{8}$ secs
which means about 10 years of storage (1 year $\sim \pi  \times 10^{7}$ secs).

\subsection{Thermal average of the hysteresis loop}
Thermal fluctuations induce random orientations of a Stoner particle.
If the change of orientation is fast with respect to our appreciation of the
hysteresis loop, then we observe an overall mean behaviour stemming from an average hysteresis loop.
This average hysteresis loop can be calculated with several methods. 
In ensemble averaging, one considers a single grain in many orientational configurations  
that is making different angles with the magnetic field (taken along the z-direction).
With time averaging, one considers a single grain undergoing different magnetization 
cycles while the magnetic field is making different angles with the grain axis.
Under the Ergodic hypothesis (see ref.~\cite{Reif}) these averaging techniques should yield the same result.
Adopting the ensemble average, we ought to find for each angle $\phi$ the 
minimum energy angle $\theta$ and every point on the hysteresis loop is made from
the average over values of $\phi$.
We perform the averaging in 3D following the original work of Stoner-Wohlfarth despite
the fact our previous description was intentionally limited to 2D. 

\begin{figure}[!ht]
\begin{center}
\scalebox{0.5}{\includegraphics*[angle=0]{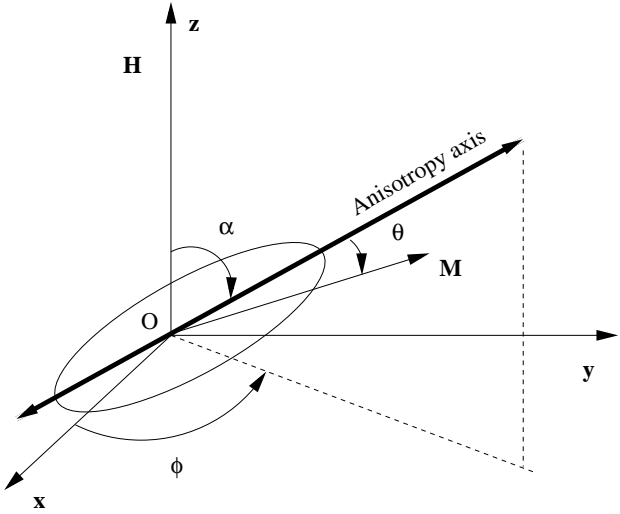}}
\end{center}
  \caption{System of coordinates displaying the anisotropy axis in 3D with the applied magnetic field $\bm{H}$ 
along the $\bm{z}$-axis and the magnetization $\bm{M}$ all in the same vertical plane indicated by dashed lines
and making the angle $\phi$ with the $\bm{xOz}$ plane.}
\label{coord}
\end{figure}

Taking the anisotropy axis along the grain long axis with polar angle $\alpha$ and
azimuthal angle $\phi$ (see fig.~\ref{coord}) let $p(\alpha,\phi)$ denote the
PDF (probability density function) of the angles $\alpha,\phi$ be uniform over the domains $[0,\frac{\pi}{2}]$ 
(see note~\cite{symmetry})
 and $[0,2\pi]$.
Hence, the average loop (being the projection of the magnetization over the direction of
the field) is given by:

\begin{equation}
\overline{\cos(\theta+\alpha)}= \frac{\int_{0}^{2\pi} d\phi \int_{0}^{\frac{\pi}{2}} \sin(\alpha) d\alpha \hspace{1mm} \cos(\theta+\alpha) p(\alpha,\phi)  }
                              {\int_{0}^{2\pi} d\phi \int_{0}^{\frac{\pi}{2}} \sin(\alpha) d\alpha \hspace{1mm} p(\alpha,\phi)  }
\end{equation}

\begin{figure}[!ht]
\begin{center}
\scalebox{0.5}{\includegraphics*[angle=0]{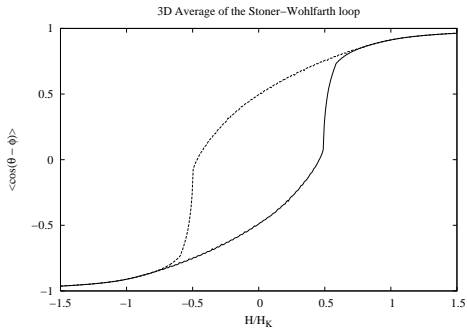}}
\end{center}
  \caption{The 3D averaged hysteresis loop looks very much like the Stoner-Wohlfarth curve except
it is less rounded at the approximate switching field values.}
\label{average}
\end{figure}

Since the individual PDF are independent, the joint PDF: $p(\alpha,\phi)=p_{\alpha}(\alpha)p_{\phi}(\phi)$
is decoupled and since both PDF are flat, we get:
\begin{equation}
\overline{\cos(\theta+\alpha)}= \int_0^{\pi/2} \cos(\theta+\alpha) \sin(\alpha) d\alpha
\end{equation}
The algorithm is now clear: Sweeping over $\alpha$ we
find the angle $\theta$ that minimises the energy in order to perform the integral.
In order to optimize the number of arithmetic operations, we rather do the following.
We transform the minimum equation (as done in the first part) in the form:
$\sin(\theta)\cos(\theta)+h \sin(\theta+\alpha)=0$ through the replacement:
$m=\cos(\theta+\alpha)$ obtaining the equation:
\begin{equation}
h_{\uparrow, \downarrow}=-m\cos(2\alpha) \pm \frac{(2m^2-1)}{2\sqrt{1-m^2}}\sin(2\alpha)
\end{equation}
with the plus sign for the upper branch and the minus sign for the lower branch.
Sweeping over values of $m$ since $|m| < 1$ allows us to find the corresponding
values of $h$ from which we keep only the minima energy values satisfying the equation:
$ \cos(2\theta)+h \cos(\theta+\alpha) \ge 0$. This gives us a table that with proper bookkeeping
will help us find the average loop.
The result of the averaging is displayed in the fig.~\ref{average} and compared in detail (see the figure caption)
to the SW work.

\subsection{Langevin dynamics for the L-L-G equations}

At finite temperature, the deterministic L-L-G equation is replaced by the
stochastic Langevin equation \cite{Reif} governing the evolution
of $\bm{M}$. The effect of temperature is contained in a random additional field  $\bm{\eta}$
(stemming from thermal white noise) acting on  $\bm{M}$: 

\begin{eqnarray}
\frac{d\bm{M}}{dt} =  -\gamma_0 (\bm{M}\times [\bm{H_e} + \bm{\eta}])  \nonumber \\
+\frac{\lambda \gamma_0}{\|\bm{M}\|} \bm{M} \times (\bm{M} \times \bm{H_e})
\label{eq:Langevin}
\end{eqnarray}

where the additional magnetic field $\bm{\eta}=(\eta_x, \eta_y,\eta_z)$ is defined by:
\begin{equation}
<\eta_i>=0,   \hspace{0.3cm} <\eta_i(t) \eta_j(t')>= 2 \Delta \delta_{ij}\delta(t-t') 
\end{equation}

where $\Delta$ is the white noise intensity given by~\cite{Brown} $\Delta=\lambda k_B T/\gamma_0 M_s$
and $(i,j=x,y,z)$.

Let ($\theta,\phi$) be the spherical angles of the orientation of the moment $\bm{M}$.
One may view ($\theta,\phi$) as a point on the surface of the unit sphere. A statistical
ensemble of moments with different orientations  can be represented by a distribution
of points over the unit sphere $W(\theta,\phi,t)$ at time $t$. Conservation of probability
leads to a continuity equation:
\begin{equation}
\frac{\partial W}{\partial t}+\nabla \cdot \bm{J}=0
\end{equation}

similar to electric charge continuity  equation. The above is in fact
a Fokker-Planck (F-P) partial differential equation (PDE) as shown below.
The (probability) current density definition $\bm{J}=W \bm{v}$ uses the velocity
 $\bm{v}=\frac{1}{M_s}\frac{d \bm{M}}{dt}$
of the point ($\theta,\phi$) on the sphere, whereas $W$ plays the role of a charge density.
 
Let us specialize to the case of a single angular degree of freedom and
apply standard methods,\cite{Gardiner} to write the PDE for the conditional probability
density $P \equiv P(x',t|x,0)$. The latter expresses the probability density
 of observing $x'=\theta(\mbox{  at time  } t)$ given the initial state 
$x=\psi(\mbox{   at time  }t=0)$. We get the following F-P equation: 
\begin{equation}
       \frac{\partial P}{\partial t}=  A(x) \frac{\partial P}{\partial x}  
    + \frac{1}{2} B(x) \frac{\partial^2 P}{\partial x^2} 
       \label{backward}
\end{equation}

The "mean first passage time"  (time, for the Stoner particle, to switch) $T(x)$
satisfies an ODE given by (see ref.~\cite{Gardiner}):

\begin{equation}
     A(x) \frac{d T(x)}{d x}
         + \frac{1}{2} B(x) \frac{d^2 T(x)}{d x^2}=-1
       \label{time}
\end{equation}

In our case, $x=\psi, A(\psi)=\frac{\lambda\gamma_0}{M_s}\frac{dE(\psi)}{d\psi}-\frac{\gamma_0^2\Delta}{\tan(\psi)}$ 
where $E(\psi)$ is the energy (per unit volume) of the Stoner particle, $E(\psi)= -M_s H \cos(\psi)+K_{eff} \sin^2(\psi)$,
containing Zeeman and effective anisotropy terms. $H$ is the externally applied field.
Besides $\frac{1}{2}B(\psi)=-\gamma_0^2 \Delta$ yield the equation for 
the "mean first passage time" as:

\begin{eqnarray}
 -\left[  \frac{\lambda\gamma_0}{M_s}\frac{dE(\psi)}{d\psi}-\frac{\gamma_0^2\Delta}{\tan(\psi)} \right] \frac{d T(\psi)}{d \psi} \nonumber \\
         + \gamma_0^2 \Delta \frac{d^2 T(\psi)}{d \psi^2}=+1
       \label{ode}
\end{eqnarray}

This second order ODE can be transformed into a first-order equation in $v(\psi)=dT/d\psi$ 
and integrated once with the initial condition  $v(\psi=0)=0$:

\begin{equation}
v(x)=\frac{1}{\sin(x)} \left[ \exp(-f(x)) \int_x^0 \frac{\sin y}{\gamma_0^2 \Delta} \exp(f(y)) dy \right]
\end{equation}

with:
\begin{equation}
 f(x)= \frac{\lambda}{4\gamma_0  V \Delta} (4HV \cos x +\beta M_s \cos 2x +\beta M_s)
\end{equation}

$V$ is the volume of the particle and $\beta=\frac{2V K_{eff}}{M_s^2}$.

Integrating once again to get $T(x)$ and using the definition of the  thermal transit 
time $t_{th}$  as the value $T(x=0)$, we obtain:

\begin{equation}
 t_{th}/c=a\int_{\cos(\theta_0)}^1 dx \frac{e^{-a{(x+b)}^2}}{(1-x^2)}
 \int_x^1 dy e^{a{(y+b)}^2}
       \label{transit}
\end{equation}

The angle $\theta_0$ maximizes the stationary PDF $\exp[-E(\theta)/k_B T] \sin(\theta)$ and is
also given by the condition $T(\theta_0)=0$. The coefficients $a,b,c$ are given respectively by:
\begin{equation} 
a=\frac{K_{eff}}{k_B T}, b=\frac{H M_s}{2 K_{eff} }, c=\frac{M_s}{ \gamma_0 \lambda  K_{eff}}
\end{equation}

It is interesting to analyze the results at high and low temperatures.
In the high temperature limit ($a \sim 0$); we get:  $t_{th} \sim ca \ln(2)$; 
whereas at low temperature ($a{(1+b)}^2 >> 1$), we obtain:

\begin{equation}
 t_{th}= \frac{c}{2} \sqrt{\frac{\pi}{a}}  \frac{1}{(1-b^2)}  \frac{1}{(1+b)} \exp[ a {(1+b)}^2] 
       \label{highT}
\end{equation}

Identifying the thermal transit time with $\tau$ we recover in that way the N\'eel-Arrhenius expression:
\begin{eqnarray} 
 t_{th} \sim \tau & =\tau_0 \exp(\Delta E/k_B T), \mbox{ where the prefactor} \nonumber \\
\tau_0  & = \frac{c}{2} \sqrt{\frac{\pi}{a}}  \frac{1}{(1-b^2)}  \frac{1}{(1+b)}, 
\end{eqnarray}

and the barrier height  $\Delta E =  K_{eff} {(1+b)}^2$ at low temperatures.

\begin{figure}[!ht]
\begin{center}
\scalebox{0.6}{\includegraphics*[angle=0]{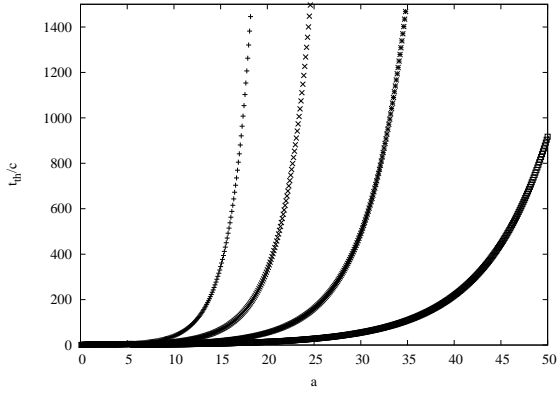}}
\end{center}
  \caption{Normalized thermal switching time $t_{th}/c$ versus inverse temperature
 $a$ for various field-anisotropy
ratios  $b=-0.3, -0.4, -0.5, -0.6$ as we proceed from left to right.}
\label{th_time}
\end{figure}

In the case of arbitrary temperature, the behaviour of the transit time versus $a$ (inverse temperature) 
for various field-anisotropy  ratios $b$ is obtained numerically as depicted in fig.~\ref{th_time}.

One might be tempted to define the thermal switching time directly from the behaviour of the
probability versus time since the F-P equation provide a means to obtain that behaviour. Switching
is reached when the probability $P(\pi/2,t_s)=0.5$.

\begin{figure}[!ht]
\begin{center}
\scalebox{0.8}{\includegraphics*[angle=0]{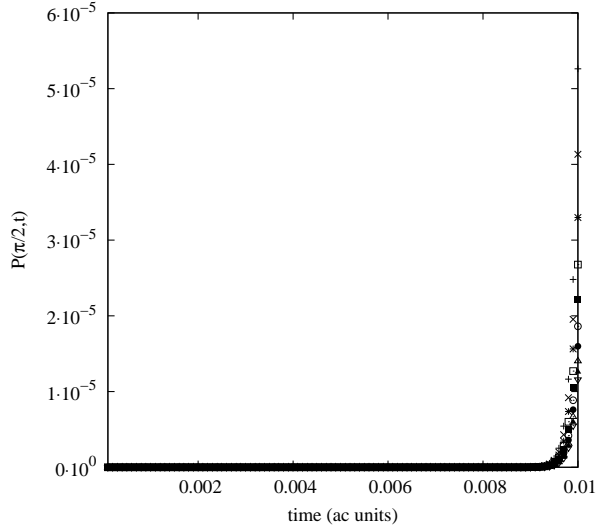}}
\end{center}
  \caption{Switching probability at $\theta=\pi/2$ versus time at a fixed 
temperature for a field-anisotropy ratio $b=-0.4$ 
(see ref.~\cite{denisov}). The ten curves corresponding to ten different inverse temperatures uniformly
distributed  over the interval $a= [0-50]$ (see fig.~\ref{th_time}) are indistinguishable. 
Switching time is reached when the probability is equal to 1/2.}
\label{prob}
\end{figure}

In fig.~\ref{prob}, the time dependence of the probability $P(\pi/2,t)$ is displayed
for the field-anisotropy ratio $b=-0.4$ (see ref.~\cite{denisov}) and shows a very quick variation above some
threshold time if one starts initially from all zero values of the probability.

\begin{figure}[!ht]
\begin{center}
\scalebox{0.8}{\includegraphics*[angle=0]{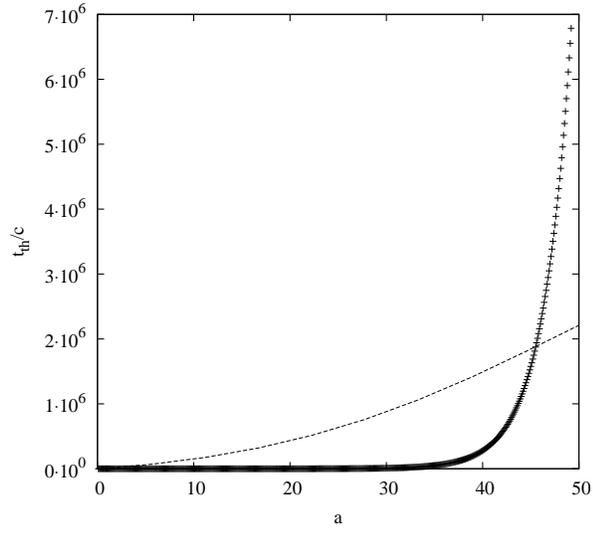}}
\end{center}
  \caption{Comparison between the analytical formula for the normalized thermal switching time $t_{th}/c$ and the numerical
integration of the F-P equation versus inverse temperature $a$. The analytical formula leads 
to a very steep variation with $a$ in sharp contrast with the numerical case.}
\label{comp}
\end{figure}

The results are validated by comparison with the analytical case in fig.~\ref{comp}.
Even if the steepness of the numerical results appear to be weaker than the anatytical
results, the graph provides a strong support for the approximate equivalence of both descriptions.

\begin{figure}[!ht]
\begin{center}
\scalebox{0.8}{\includegraphics*[angle=0]{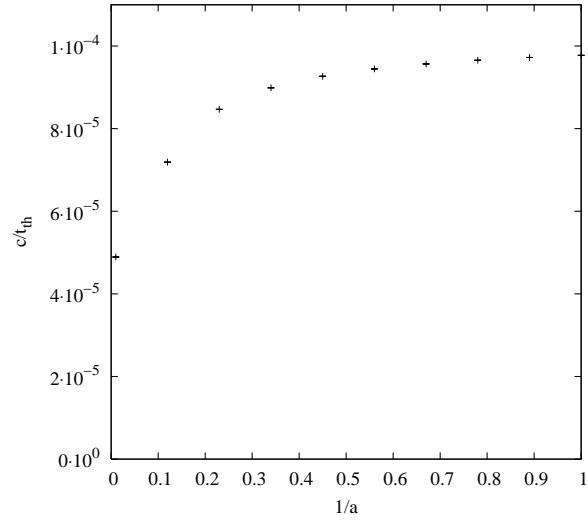}}
\end{center}
  \caption{Inverse normalized thermal switching time $c/t_{th}$ versus temperature $1/a$ for a field-anisotropy ratio $b=-0.4$.}
\label{invts}
\end{figure}

Analysis of the thermal switching time versus temperature has numerous technological consequences.
Once again, the F-P equation provides this kind of information paving the way
to the search of the best materials/conditions that yield the optimal switching time. \\
We perform direct time integration of the F-P 
equations to extract the behaviour of the switching time
versus temperature. As an illustration, using the same field-anisotropy  ratio  
$b=-0.4$ as previously, the inverse switching time versus temperature 
is displayed in fig.~\ref{invts}.

\section{Extensions and perspectives of the Stoner-Wohlfarth model}
The SW model is a macrospin description of magnetic systems that is extremely rich from 
the static, dynamic and statistical viewpoints. 
Despite its numerous limitations (and of the macrospin approach in general)
described in the first part of this work, it remains
a valid starting point for the useful description and basic understanding 
of many (static and dynamic) problems of fundamental and applied magnetism.

The full 3D counterpart of the SW model as done in ref.~\cite{wern} is an important extension and of great interest. 
We point anew to the fact the averaging of the hysteresis loop done 
in section 3 related to thermal effects was performed in 3D as in SW work for comparison.

The extension to uniaxial anisotropies of arbitrary order (a higher anisotropy is
of fourth or sixth order like in Cobalt ...) or other forms like biaxial, planar,
cubic (as in solid Ni or Fe) or of several competing types might provide a richer behaviour of the 
loop versus angle.

The use of arbitrary non-ellipsoidal shape for the grain is also challenging given
the occurrence of non-uniformity of the magnetization. \\ 

The interaction between grains must also be studied and gauged with respect to
its role in affecting the switching of the magnetization. 
New types of interactions or novel types of exchange between grains or with other objects
might be exploited in spintronic and quantum devices.
 
{\textbf Acknowledgement} \\
The  authors wish to acknowledge W. D. Doyle (MINT, Alabama) for sending many papers 
of his work on fast switching and helpful correspondance. \\

\begin{center}
{\textbf FIGURES}
\end{center}

\begin{thebibliography}{99}
\bibitem{symmetry} The polar angle $\alpha$ varies in general over the interval $[0,\pi]$. From symmetry it is
restricted here to the interval $[0,\frac{\pi}{2}]$.
\end{thebibliography}
\end{document}